\documentclass[aps,prd,twocolumn,eqsecnum,showpacs,preprintnumbers,amsmath,amssymb]{revtex4-1}
\usepackage{graphicx}
\usepackage[usenames]{color}
\DeclareMathOperator{\Tr}{Tr}
\DeclareMathOperator{\tr}{tr}
\begin{document}
\preprint{FERMILAB-PUB-13-280-T}
\def\beq{\begin{equation}}
\def\enq{\end{equation}}
\title{Exact blocking formulas for spin and gauge models}
\author{Yuzhi Liu$^{1, 2}$}
\author{Y. Meurice$^1$}
\author{M. P. Qin$^3$}
\author{J. Unmuth-Yockey$^1$}
\author{T. Xiang$^{3}$}
\author{Z. Y. Xie$^3$}
\author{J. F. Yu$^3$}
\author{Haiyuan Zou$^1$}
\affiliation{$^1$ Department of Physics and Astronomy, The University of Iowa, Iowa City, Iowa 52242, USA }
\affiliation{$^2$ Theoretical Physics Department, Fermi National Accelerator Laboratory, Batavia, Illinois 60510, USA}
\affiliation{$^3$ Institute of Physics, Chinese Academy of Sciences, P.O. Box 603, Beijing 100190, China}

\def\lt{\lambda ^t}
\def\note{note}

\date{\today}
\begin{abstract}
Using the example of the two-dimensional (2D) Ising model, we show that in contrast to what can be done in configuration space,  the tensor renormalization group (TRG) formulation allows one to write exact, compact, and manifestly local blocking formulas and exact coarse grained expressions for the partition function. We argue that similar results should hold for most models studied by lattice gauge theorists. We provide exact blocking formulas for several 2D spin models (the $O(2)$ and $O(3)$ sigma models and the $SU(2)$ principal chiral model) and for the 3D gauge theories with groups $Z_2$, $U(1)$ and $SU(2)$. We briefly discuss generalizations to other groups, higher dimensions and 
practical implementations. 
\end{abstract}


\maketitle

\section{Introduction}
    Lattice field theory is a well-developed numerical method which allows us to study the nonperturbative behavior of asymptotically free theories. 
    The continuum limit of these theories is reached in the limit of zero bare coupling. If a mass gap remains present in this limit, the lattice spacing becomes exponentially small 
    compared to the physical scale associated with the mass gap. Keeping the physical volume reasonably large requires exponentially large volume or clever extrapolations. 
    For gauge theories with enough massless fermions, a nontrivial infrared fixed point may appear (for a review see Ref. \cite{DeGrand:2010ba}). Evidence for quasi-conformal behavior requires small masses and large volume. 
    The practical demands of these two limits are sometimes too taxing to lift controversies regarding the existence of a nontrivial fixed points.  A well-known example is $SU(3)$ with 12 fundamental fermions \cite{Fodor:2011tu,Deuzeman:2012ee,Cheng:2011ic,Aoki:2012eq,Appelquist:2011dp} .

    The above considerations make clear that reaching exponentially large volumes is a highly desirable outcome for lattice field theorists. 
    This goal could be reached if a reasonably accurate blocking procedure could be designed. By blocking (or block-spinning or coarse-graining) 
    we mean a partial integration procedure used in the renormalization group (RG) approach \cite{wilson74,cardy96,Meurice:2011wy} to replace the initial degrees of freedom on sites or links corresponding to a lattice spacing $a$ by some new ones assigned to the sites or links corresponding to the a lattice spacing $ba$ while keeping the macroscopic observables and extensive quantities unchanged. We call $b$ the scaling factor (typically $b=2$).  By blocking $n$ times, linear lattices of order $b^n$ can be reached which is the aforementioned goal. 

    It is often believed that blocking in configuration space, for instance by summing over the spins in a block while keeping their sum constant, is tedious but straightforward. The example of the 2D Ising model with two-by-two blocks (briefly discussed in Sec. \ref{sec:ising}) can be used to show that the procedure is far from straightforward because it generates an arbitrarily large number of new interactions of arbitrarily large range that are difficult to enumerate and control. 
    It is possible to invent approximations where no new interactions are generated by the blocking process. Examples are the Migdal-Kadanoff approximation \cite{Migdal:1975zf,Kadanoff:1976jb}, the approximate recursion formula \cite{wilson71b}  or other hierarchical approximations \cite{baker72,hmreview}. However, the lack of reference to an exact procedure makes the systematic improvement of these approximations difficult. 

    In this paper, we show that, in contrast to the difficult situation encountered in configuration space, the tensor renormalization group (TRG) formulation allows us to write {\it exact} blocking formulas for several classes of spin and gauge models. 
    For these models, the partition function can be written as a product of tensors attached to sites (or links, or plaquettes) with their indices suitably contracted (traced). After blocking, the partition function has exactly the same form as before except for the fact that the lattice spacing is twice bigger and that the sum over the indices has more terms. 
    The recursion formulas for the tensors are manifestly local and do not generate new types of tensors. However, the ability to reinterpret these results in terms of blocked configurations and interactions is probably lost. 

    The TRG approach of classical lattice models was introduced in Refs. \cite{JPSJ.64.3598,PhysRevLett.99.120601,PhysRevB.79.085118} and was motivated by 
    tensor states developed in RG studies of quantum models \cite{uli}. 
    For this reason, we often refer to sums over tensor indices as sums over states. 
    Improved methods to take into account the environment were proposed in Refs.  \cite{PhysRevLett.103.160601,PhysRevB.81.174411}. One important purpose of our article is to show that TRG methods can be applied to many models studied by lattice gauge theorists and that detailed comparisons with standard Monte Carlo simulations should be performed. 

   In Sec. \ref{sec:ising}, we start with the well-understood case of the 
    2D Ising model on a square lattice for which very accurate TRG based numerical calculations \cite{PhysRevB.86.045139} were performed. 
    The construction of the initial tensor can be performed using singular value decomposition (SVD). This task can be simplified by using the character expansion 
    techniques used to reformulate lattice models in terms of dual variables \cite{RevModPhys.52.453}. As the recursion formula is iterated, the number of states increases rapidly and truncation  methods are necessary. Optimal methods with apparent convergence when going to a sufficiently large number of states were discussed in \cite{PhysRevB.86.045139}. It is interesting to notice that 
    two states approximations provide much better estimates of the critical exponents \cite{PhysRevB.87.064422} than the Migdal-Kadanoff approximation. 
    However, there is an intermediate region for the number of states retained where oscillations appear in individual tensor components and new techniques need to be developed if we want to work within this intermediate number of states zone. This situation has been recently documented and analyzed in Ref. \cite{efrati13} which also provides a nice introduction to the TRG method. 

     The rest of the paper is organized as follows. We provide the exact tensor recursion formulas for the 2D $O(2)$ and $O(3)$ sigma models (Sec. \ref{sec:spin}), the 2D $SU(2)$ principal chiral model (Sec. \ref{sec:chiral}) and the 3D 
    $Z_2$, $U(1)$ and $SU(2)$ pure gauge theories (Sec. \ref{sec:gauge}). Again, the calculation of the initial tensor can be done from expansions used in dual formulations 
    \cite{PhysRevD.17.2637,Orland:1979rh,RevModPhys.52.453,PhysRevD.30.1775,Anishetty:1992xa} even though we do not deal with the dual variables here. The generalizations of these results and ongoing practical applications are discussed in the conclusions.

\section{Blocking the 2D Ising model}
    \label{sec:ising}

    In order to appreciate the importance of having exact blocking formulas, we discuss the 2D Ising model on a square lattice with nearest neighbor ferromagnetic interactions and an inverse temperature $\beta$. We first work in configuration space (with spin and blocked spin variables) and then 
    with the TRG approach. Some of the features observed generalize to other Abelian spin models. 

    \subsection{Configuration space blocking}

\begin{center}
            \begin{figure}[h]%
            \includegraphics[width=3in]{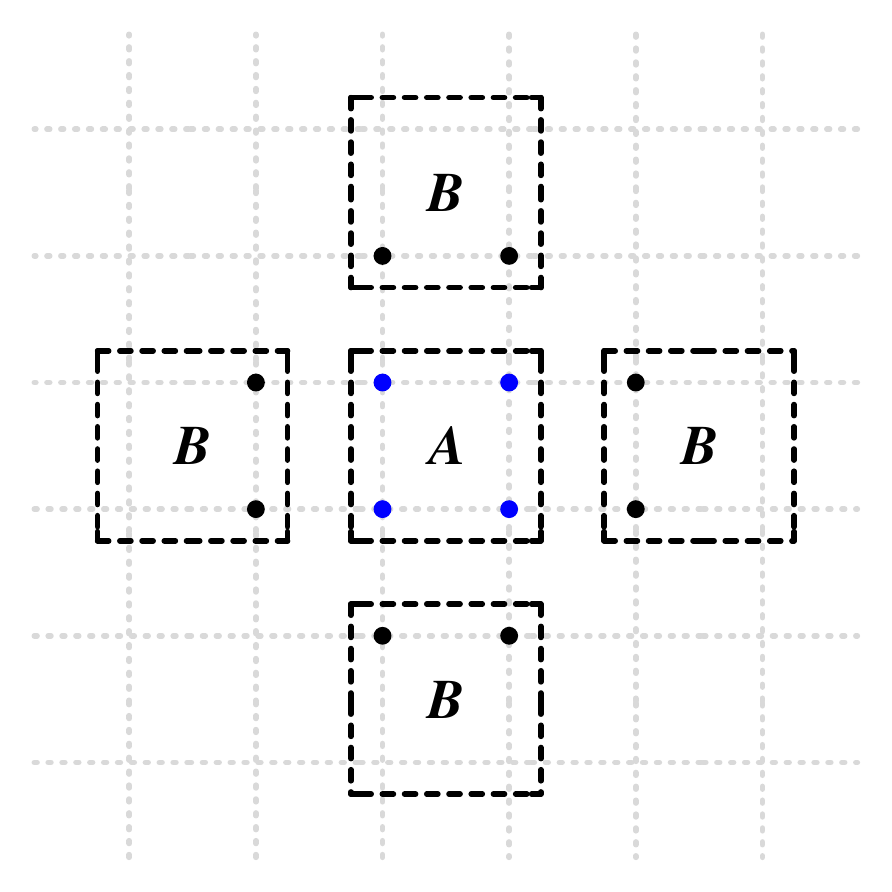}%
            \caption{AB checkerborad}%
            \label{fig:AB0}%
            \end{figure}
        \end{center}

\begin{center}
            \begin{figure}[h]%
            \includegraphics[width=3in]{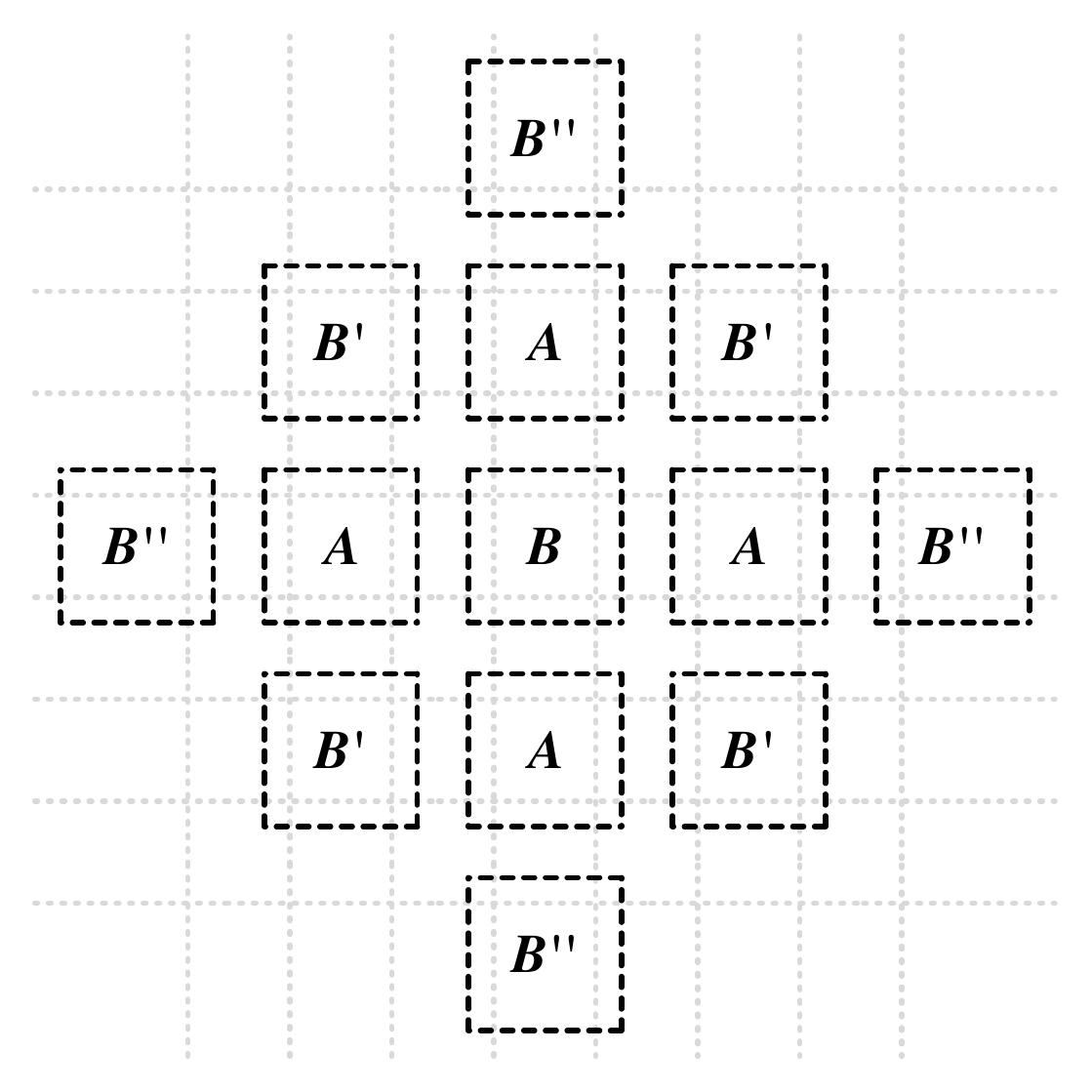}%
            \caption{AB checkerborad}%
            \label{fig:AB}%
            \end{figure}
        \end{center}

        We first describe an algorithm to block spin {\it once} in configuration space and try to generate a new energy function. The main purpose of the discussion is to show that unless approximations are made, this new 
        energy function is very different from the original one. We consider square blocks with a checker board configuration partitioning the blocks into 
        $A$ and $B$ blocks as illustrated in Figs. \ref{fig:AB0} and \ref{fig:AB}. 
        We treat the spins in the $B$ blocks as fixed background and proceed to calculate unnormalized probabilities for the total spin in the $A$ blocks. For a given $A$ block, there are 8 background spins belonging to the  four nearest neighbor $B$ blocks as shown in Fig. \ref{fig:AB0}. 
        The total spin $\phi_A$  in the $A$ block takes values $\pm 4$, $\pm 2$ and 0. For each value of $\phi_A$ and for each of the background configurations, 
        we can sum over the known Boltzmann weights and obtain $5\times 2^8$ unnormalized probabilities. 

        The next step is to try to block spin in the $B$ blocks. We consider a given $B$ block as the one at the center of Fig. \ref{fig:AB}. 
        This can be done by combining our previous results for the four nearest neighbor $A$ blocks. We can now sum over the Boltzmann weights corresponding to each of the values $\pm 4$, $\pm 2$ and 0 in the $B$ block. 
        This clearly generates probabilities for the $5^5$ configurations of the four $\phi_A$s and the central $\phi_B$ and consequently generates nearest neighbor interactions among these blocked variables. 
        The crucial point is that the results depend on the remaining background spins attached to the four $A$ blocks and so the sums over the remaining spins in the $B$ blocks cannot be done independently of each others. 
        There are twelve such spins in the four $B$ blocks.  Denote these four blocks $B'$ that are located diagonally from the central $B$ block under consideration.  There are eight spins in the next to nearest neighbors $B$ blocks denoted $B''$. 
        If we try to construct the new energy function numerically, this requires $5^5\ 2^{20}$ (about 3 billions) memory entries. 
        It is clear that these correlations will generate more than nearest neighbor interactions. 

        We should now pause and discuss what has been accomplished so far and what remains to be done. For this purpose, we can tile the original lattice with the ``diamonds" of Fig. \ref{fig:AB}.
        The unshared spins inside the diamonds have been blocked. This is 5/8 of the total number of spins. The remaining spins (3/8 of the total number of spins) are at the shared boundaries of the diamonds and form diagonals. 
        The $B'$ blocks are shared by two diamonds and the $B''$ blocks are shared by 4. 
        One can in principle combine  4 such diamonds into a new diamond with twice the linear size. This discussion makes clear that arbitrary range interactions are generated and finding a new energy 
        function in terms of polynomials of the $\phi_A$ and $\phi_B$  seems to be a Herculean task at least as difficult as calculating the 
        exact partition function in a comparable volume. 

    \subsection{TRG blocking}

        In contrast, blocking is amazingly simple in the TRG formulation. For each link we can write 
        \begin{eqnarray}
        \exp(\beta \sigma_1 \sigma _2)&=&\cosh(\beta)(1+\sqrt{\tanh(\beta)}\sigma_1    \sqrt{\tanh(\beta)}\sigma_2 )    \nonumber \\
        =\cosh(\beta) &\sum& _{n_{12}=0,1} (\sqrt{\tanh(\beta)}\sigma_1    \sqrt{\tanh(\beta)}\sigma_2)^{n_{12}} .
        \label{eq:char}
        \end{eqnarray}
        Using this identity for each link in the partition function, we can then regroup the four terms involving a given spin $\sigma_i$ and sum over its two values $\pm 1$. The results can be expressed in terms of a tensor 
        $T^{(i)}_{xx'yy'}$ which can be visualized as a cross attached to the site $i$ with the four legs covering half of the four links attached to $i$. 
        The horizontal indices $x,\ x'$ and vertical indices $y,\ y'$ take the values 0 and 1 as the index $n_{12}$ in Eq. (\ref{eq:char}). The tensor is zero for an odd number of 1s. For an even number of 1s, a factor 
       $ \sqrt{ \tanh(\beta})$ appears for each 1 irrespective of the direction.  This can be summarized as follows:
                       \begin{equation}
            T^{(i)}_{xx'yy'} =f_x f_{x'}f_y f_{y'} \delta\left(\text{mod}[x+x'+y+y',2]\right) \ ,
            \label{eq:factor}
        \end{equation}
        where $f_0=1$ and $f_1 =\sqrt{ \tanh(\beta)}$. The delta symbol is 1 if $x+x'+y+y'$ is zero modulo 2 and zero otherwise. 
              
        The partition function of the model can now be written as 
        \begin{equation}
            Z = \Tr \prod_{i}T^{(i)}_{xx'yy'} \ .
            \label{eq:Z}
        \end{equation}
         $\Tr$ 
        is a short notation for contractions (sums over 0 and 1) over the links joining nearest neighbors on the lattice. This expression reproduces the proper closed paths of the high-temperature expansion.  
        
\begin{figure}[h]
            \includegraphics[width=2.3in,angle=0]{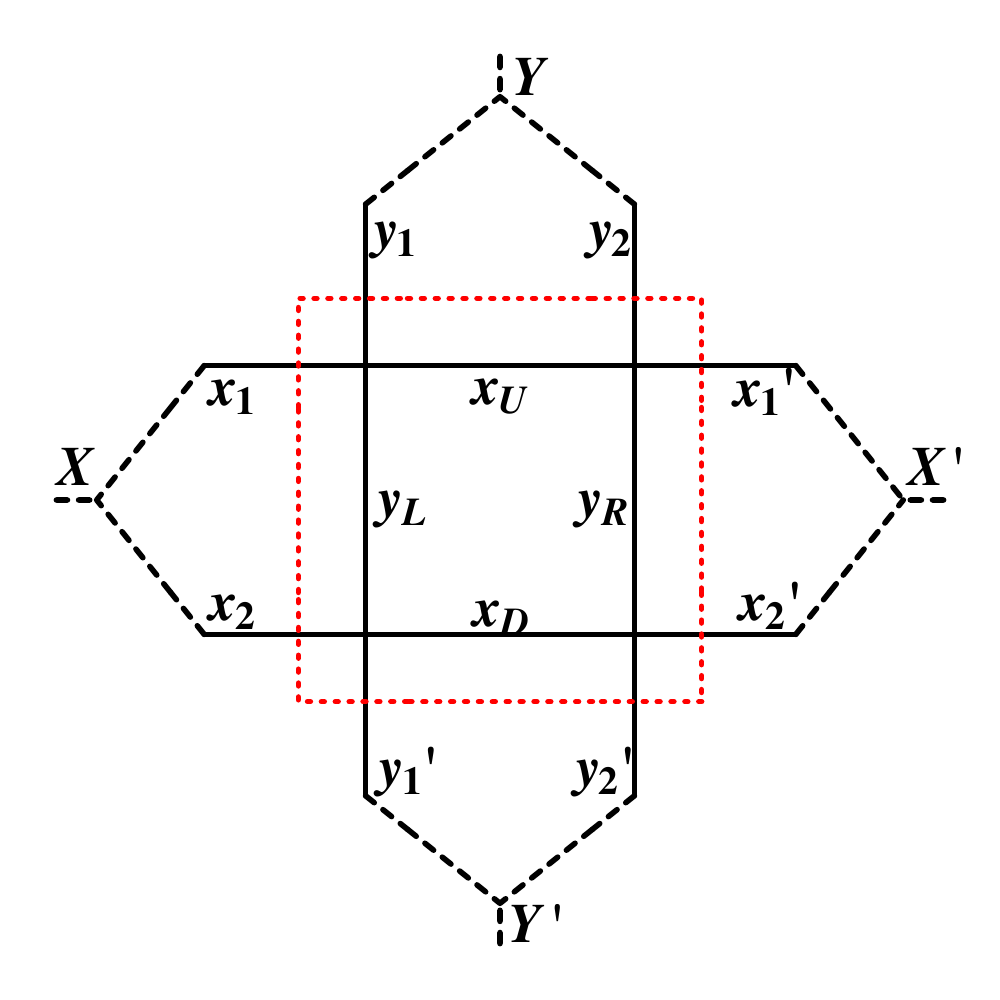}
            \caption{Graphical representation of $T'_{XX'YY'}$ . 
            \label{fig:square} }
        \end{figure}

        We now use this reformulation to blockspin \cite{PhysRevB.86.045139,PhysRevB.87.064422}. 
        We consider an  isotropic procedure with a square block 
        enclosing four sites as in the previous subsection and sum over the states inside the block associated with the nearest neighbor links joining these four points. This defines a 
        new rank-4 tensor $T'_{XX'YY'}$ where each index now takes four values. 
        \begin{eqnarray}
            \label{eq:square}
            &\ &T'_{X({x_1},{x_2})X'(x_1',x_2')Y(y_1,y_2)Y'(y_1',y_2')} = \\ \nonumber
            &\ &\sum_{x_U,x_D,x_R,x_L}T_{x_1 x_U y_1y_L}T_{x_Ux_1'y_2y_R}T_{x_Dx_2'y_R y_2'}T_{x_2x_Dy_Ly_1'}\  ,
        \end{eqnarray}
        where $X(x_2,x_2)$ is a notation for the product states. In Ref.  \cite{PhysRevB.87.064422} we used the convention: 
        $X(0,0)=1,\  X(1,1)=2, \  X(1,0)=3,\  X(0,1)=4$. 
        This is represented graphically 
        in Fig.  \ref{fig:square}. 
        The partition function can be written as 
        \begin{equation}
            Z=\Tr\prod_{2i}T'^{(2i)}_{XX'YY'} \ , 
            \label{eq:ZP}
        \end{equation}
        where $2i$ denotes the sites of the coarser lattice with twice the lattice spacing of the original lattice. 
        As pointed out in Ref. \cite{PhysRevB.87.064422} the TRG blocking is exact and can be written compactly because the procedure separates, unambiguously, the degrees of freedom inside the block which are integrated over from those kept to communicate with the neighboring blocks.

    \subsection{Abelian factorization and connection to configuration space}
    \label{subsec:factor}
         Eq. (\ref{eq:factor}) shows that the initial tensor factors nicely.  This property can be extended to spin models with an Abelian group. For an explicit example see $O(2)$ in Sec. \ref{subsec:o2}. The general reasoning goes as follows. The Boltzmann weight associated with a link can be expanded in characters (Fourier modes). For an Abelian group, each character of the expansion is a product of two characters involving each of the site variables. Similarly the coefficient of the expansion can be written as the product of two square roots of itself, each one being associated with a given site. One can then regroup all the characters and square roots associated with the variable of a given site. The initial tensor is obtained after integrating over the site variable. This gives a Kronecker delta times a product similar to what is seen in Eq. (\ref{eq:factor}). The reasoning immediately extends to arbitrary dimension.

        The sum over the internal states in Fig. \ref{fig:square}  and Eq. (\ref{eq:square}) is very similar to a sum over momenta in Feynman diagrams. Three of the sums are absorbed using the Kronecker delta associated with three vertices, but there is a global condition on the external legs that, if satisfied, guarantees that the fourth condition is satisfied. Consequently there is, in general, 
        a sum over ``states circulating in the loop" which is enclosed in the block. This sum of factorizable terms is apparently not factorizable and it seems impossible to rewrite 
        the blocked tensor as coming from a blocked energy function with an Abelian symmetry. This argument indicates that a direct connection between blocked tensors and blocked energy functions might be difficult or impossible to find.

\section{TRG formulation of $O(2)$ and $O(3)$ sigma models}
    \label{sec:spin}

    The Hamiltonian for the $O(N)$ nonlinear sigma models can be written as
    \beq
    H = - \sum_{<ij>}{\mathbf{S}_\mathbf{i}\cdot \mathbf{S}_\mathbf{j}}\ ,
    \enq
    with $\mathbf{S}_\mathbf{i}$ a unit vector in $\mathbb{R}^N$, or equivalently a point on a $N$-dimensional unit sphere. 
    We will discuss explicitly the Abelian case $N=2$ and the non-Abelian case $N=3$ in two dimensions. In both cases, the TRG expression of the partition function has  
    the same form as Eq. (\ref{eq:Z}) for the Ising model. The only difference being the range of the indices and the initial values. Similarly, the blocking of the tensor has the same form as Eq. (\ref{eq:square}) and will not be written explicitly. 

    \subsection{$O(2)$ model}
        \label{subsec:o2}

        For $N=2$,  $\mathbf{S}_\mathbf{i}$ is a unit vector staying at each site $i$: ($\cos(\theta_i)$, $\sin(\theta_i)$).
        The partition function reads
        \beq
            Z = \int{\prod_i{\frac{d\theta_i}{2\pi}} {\rm e}^{\beta \sum\limits_{<ij>} \cos(\theta_i - \theta_j)}}.
            \label{eq:bessel}
        \enq
        Using
        \beq
            {\rm e}^{\beta  \cos(\theta_i - \theta_j)} = \sum\limits_{n_{ij}=-\infty}^{+\infty} {\rm e}^{i n_{ij}(\theta_i-\theta_j)} I_{n_{ij}}(\beta)\  ,
        \enq
        where the $I_n$ are the modified Bessel functions of the first kind. 
        From the basic property of the exponential, it is possible to collect all the factors involving a given $\theta_i$ and integrate over this variable.  
        This results in a tensor attached to the site $i$. In two dimensions, 
        \begin{eqnarray}
            T^i_{n_{ix},n_{ix'},n_{iy},n_{iy'}} &=& \sqrt{I_{n_{ix}}(\beta)} \sqrt{I_{n_{iy}}(\beta)} \sqrt{I_{n_{ix'}}(\beta)} \sqrt{I_{n_{iy'}}(\beta)}\nonumber \\
            & \ & \delta_{n_{ix}+n_{iy},n_{ix'}+n_{iy'}} \ .
        \end{eqnarray}
        The sign convention is that we have positive signs for the left and top indices and negative signs for the right and bottom indices. 
        This allows us to write the partition function and the blocking of the tensor similar to the Ising model.
        The only difference is that the sums run over the integers. As the $I_n (\beta)$ decay rapidly for large $n$ and fixed $\beta$ (namely like $1/n!$) there is no convergence issue. 
        The generalization to higher dimensions is straightforward ($2D$ indices in $D$ dimensions).

    \subsection{$O(3)$ model}
\begin{center}
            \begin{figure}[!ht]%
                \includegraphics[width=3in]{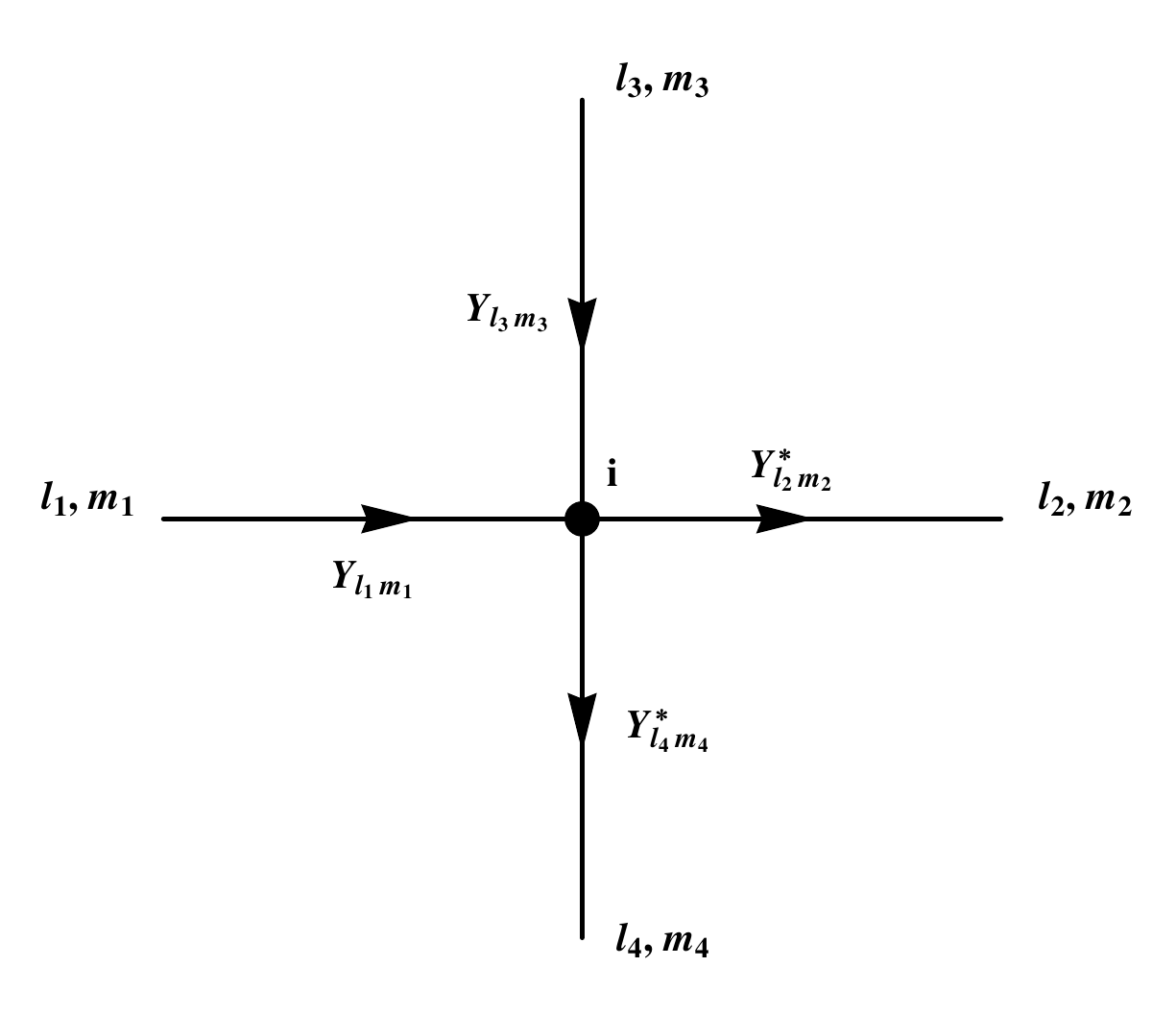}%
                \caption{2D $O(3)$}%
                \label{fig:ylm}%
            \end{figure}
        \end{center}

        For $N=3$,  $\mathbf{S}_\mathbf{i}$ is a  unit vector at site $i$: ($\sin(\theta_i) \cos(\phi_i) $, $\sin(\theta_i) \sin(\phi_i) $, $\cos(\theta_i) $). In terms of these variables, the energy function can then be written as
        \beq
            H   =   - \sum_{<ij>}{\cos\gamma_{ij}},
        \enq
        where $\gamma_{ij}$ is the angle between $\mathbf{S}_\mathbf{i}$ and $\mathbf{S}_\mathbf{j}$ and $\cos\gamma_{ij}$ can be expressed in terms of the angles         as
        \beq
            \cos\gamma_{ij}  =   \cos\theta_i \cos\theta_j + \sin\theta_i \sin\theta_j \cos(\phi_i -\phi_j) \  .
        \enq
        Expanding as for $O(2)$ and using $I_n(\beta)=I_{-n}(\beta)$, 
        \beq
            {\rm e}^{\beta \cos\gamma}=I_0(\beta) + 2\sum\limits_{n=1}^\infty I_n(\beta) \cos(n \gamma)\  ,
        \enq
        we can then use the Chebyshev polynomials of the first kind to re-express
        \beq
            \cos(n \gamma)=T_n(\cos\gamma)    \ .
            \label{eq:cheby}
        \enq
        Using the Legendre polynomials $P_n(\cos\gamma)$, we can write
        \begin{equation}
            {\rm e}^{\beta \cos\gamma_{ij}}    =   I_0(\beta) + 2\sum\limits_{n=1}^{\infty} I_n(\beta)\sum\limits_{l=0}^{n}a_{nl}P_l(\cos\gamma_{ij}), 
            \label{eq:eip}
        \end{equation}
        with
        \beq
            a_{nl}  =   \frac{2l+1}{2}\int\limits_{-1}^1 T_n(x)P_l(x)dx.
            \label{eq:anl}
        \enq
        By using the addition theorem for spherical harmonics
        \beq
            P_l(\cos\gamma_{ij}) =   \frac{4\pi}{2l+1}\sum\limits_{m=-l}^{l}Y_{lm}^*(\theta_j, \phi_j)Y_{lm}(\theta_i, \phi_i) \ ,
        \enq
        Eq. (\ref{eq:eip}) can be written as
        \beq
            {\rm e}^{\beta \cos\gamma_{ij}}    =   \sum\limits_{l=0}^{\infty} A_l(\beta) \sum\limits_{m=-l}^{l}Y_{lm}^*(\theta_j, \phi_j)Y_{lm}(\theta_i, \phi_i),
            \label{eq:ebyy}
        \enq
        where 
        \begin{align}
            A_0(\beta)  &=  I_0(\beta) + 2\sum\limits_{n=1}^{\infty} I_n(\beta) a_{n0} 4\pi ,\nonumber   \\
            A_1(\beta)  &=  2\sum\limits_{n=1}^{\infty} I_n(\beta) a_{n1} \frac{4\pi}{3}    ,        \label{eq:aaa}\\
            \vdots      &                                                               \nonumber\\
            A_l(\beta)  &=  2\sum\limits_{n=l}^{\infty} I_n(\beta) a_{nl} \frac{4\pi}{2l+1}.         \nonumber
        \end{align}
        Again, the Bessel functions control the decay for large $l$. 
        The elements of the $T$ tensor can then be written as
        \begin{eqnarray}
            &\  &T_{(l_1, m_1), (l_2, m_2), (l_3, m_3), (l_4, m_4)} = \nonumber \\
            &\  &\int_0^{2\pi} d\phi \int_0^{\pi} d\theta \sin\theta Y_{l_1 m_1}(\theta, \phi) Y^\star_{l_2 m_2}(\theta, \phi)\nonumber\\
            &\  &         Y_{l_3 m_3}(\theta, \phi) Y_{l_4 m_4}^*(\theta, \phi) \sqrt{A_{l_1} A_{l_2} A_{l_3} A_{l_4}}  .
            \label{eq:yyyy}
        \end{eqnarray}
        The direction convention is shown in Figure \ref{fig:ylm}. 
        Equation (\ref{eq:yyyy}) can be further simplified by expanding the product of two spherical harmonics in terms of spherical harmonics themselves:
        \beq
            Y_{l_1 m_1}(\theta,\phi) Y_{l_3 m_3}(\theta,\phi) = \sum\limits_{L = l_{min}}^{l_{max}} G_L^{({m_1, m_3, l_1, l_3})} Y_L^{m_1 + m_3}(\theta, \phi) \ .
            \label{eq:yygy}
        \enq
        Explicit formulas for $G_L^{({m_1, m_3, l_1, l_3})}$ and a discussion of the Gaunt coefficients can be found in \cite{Varshalovich:1988ye}.
       The summation bounds are 
        \begin{align}
            l_{max}     &=  l_1 +l_3\nonumber\\
            l_{min}     &=  \left\{ \begin{array}{lll}
                            \lambda_{min}        &\mbox{if}  &l_{max} + \lambda_{min} \mbox{ is even}\\
                            \lambda_{min}+1  &\mbox{if}     &l_{max} + \lambda_{min} \mbox{ is odd}
                            \end{array}\right. \label{eq:bounds}\\
            \lambda_{min}   &=  max(|l_1 -l_3|, |m_1+m_3|).\nonumber
        \end{align}
        The angular integration in Eq. (\ref{eq:yyyy}) can be performed using the orthonormal property of the spherical harmonics with result
         \begin{eqnarray}
            &\  &T_{(l_1, m_1), (l_2, m_2), (l_3, m_3), (l_4, m_4)} =  \delta_{m_1 + m_3, m_2 + m_4}\\
            &\ &  \sum\limits_{L} G_{L}^{({m_1, m_3, l_1, l_3})} {G_{L}^*}^{({m_2, m_4, l_2, l_4})} \sqrt{A_{l_1} A_{l_2} A_{l_3} A_{l_4}}. \nonumber 
            \label{eq:tensor}
        \end{eqnarray}
        In contrast to $O(2)$, 
        there are now two indices associated with each leg of the tensor and the factorization of the initial tensor is lost. 

\section{TRG for $SU(2)$ Principal Chiral Models}
    \label{sec:chiral}
    Using the conventions from \cite{Varshalovich:1988ye}, we start with a partition function for a Principal Chiral Model
    \begin{equation}
        Z = \prod_{n} \int dU(n) \prod_{ni} \exp{\left\{ \frac{\beta}{2} {\rm Re}[ \tr \left[ U(n)U^{\dagger}(n + i) \right] ] \right\}}.
    \end{equation}
    with $i$ a unit vector in one of the spatial directions and $n$ a spatial location.  Then since the action only depends on the trace of the matrix representation of the group elements we can write it in terms of a character expansion
    \begin{equation}
    \label{charexp}
        \exp{\left\{ \frac{\beta}{2} {\rm Re}[ \tr \left[ U(n)U^{\dagger}(n + i) \right] ] \right\}} = \sum_{r}F_{r}(\beta)\chi^{r}(U(ni)).
    \end{equation}
    with the sum over the representations of the group and $U(ni)$ a short notation for   $U(n)U^{\dagger}(n + i) $. $\chi_{r}$ is the trace in the irreducible representation $r$ of $SU(2)$.  We can rewrite the partition function as
    \begin{equation}
        Z = \prod_{n} \int dU(n) \prod_{ni} \sum_{r(ni)}F_{r(ni)}(\beta)\chi^{r(ni)}(U(ni)).
    \end{equation}
    Let $A \in SU(2)$ and let $D^{r}_{mn}(A)$ be the matrix elements in the $r^{\text{th}}$ irreducible representation (the ``Wigner D-functions'').  Then to extract the angle dependence inside of the $\chi$s we note that the $\chi$s are the traces of these representations and thus
    \begin{equation}
        \chi^{r}(AB) = D^{r}_{mn}(A)D^{r}_{nm}(B). 
    \end{equation}
    For a $2D$ lattice, there are four times when a product of pairs of sites contain the same site so if we product out all of the nearest neighbor pairs and collect the single same site together
    \begin{align}
    \nonumber
        Z &= \prod_{n} \int dU(n) \prod_{ni} \sum_{r(ni)}F_{r(ni)}(\beta) \\
        &\times \sum_{m,k}D^{r(ni)}_{mk}(U(n))D^{r(ni)}_{km}(U(n + i)) \\
    \nonumber
        &= \sum_{\{r\text{'}s\}}\sum_{\{m\text{'}s\}}\sum_{\{n\text{'}s\}} \prod_{l} \left( F_{r_{1,l}}(\beta)F_{r_{2,l}}(\beta)F_{r_{3,l}}(\beta)F_{r_{4,l}}(\beta) \right)^{\frac{1}{2}} \\ 
        & \times \prod_{n} \int dU(n) D^{r_{1,l}}_{m_{1}n_{1}}(U)D^{r_{2,l}}_{m_{2}n_{2}}(U) D^{r_{3,l}}_{m_{3}n_{3}}(U)D^{r_{4,l}}_{m_{4}n_{4}}(U) .
    \end{align}
    with $l$ a product over the sites of the lattice, and $r_1, r_2, r_3, r_4$ the four links incoming and outgoing from the site.  Each $F_r$ is shared by two sites on the lattice since they are located on the links.  Then from the integrals we get a constraint at each site on the lattice, and a product over constraints and link variables, $F_{r}$.

    To preform the integration over the site variables we can use the Clebsch-Gordon series to re-write two D-functions as a single D-function with accompanying Clebsch-Gordon symbols.
    \begin{align}
    \nonumber
        & D^{r_1}_{m_1 n_1}(U) D^{r_2}_{m_2 n_2}(U) = \\
        &\sum_{r = |r_1 - r_2|}^{r_1 + r_2} \sum_{m, n} C^{r_1 r_2 r}_{m_1 m_2 n} D^{r}_{m n}(U) C^{r_1 r_2 r}_{n_1 n_2 n}.
    \end{align}
    Then for the integrals above we can change out the four $D$s for two, and using $D^{r}_{m n} = (-1)^{n - m} {D^{r}}^{*}_{-m -n}$ and their orthogonality \cite{Anishetty:1992xa}
    \begin{equation}
        \int dU \, D^{r_1}_{m_1 n_1}(U) D^{*r_2}_{m_2 n_2}(U) = \frac{1}{2 r_1 + 1} \delta_{r_1 r_2} \delta_{m_1 m_2} \delta_{n_1 n_2},
    \end{equation}
    we can write down the integral exactly:
    \begin{align}
    \nonumber
        &\int dU \; D^{r_{1}}_{m_{1}n_{1}}(U)D^{r_{2}}_{m_{2}n_{2}}(U) D^{r_{3}}_{m_{3}n_{3}}(U)D^{r_{4}}_{m_{4}n_{4}}(U) \\
    \nonumber
        & = \sum_{r', m', n'} \sum_{r'', m'', n''} C^{r_1 \, r_2 \, r'}_{m_1 \, m_2 \, m'}C^{r_1 \, r_2 \, r'}_{n_1 \, n_2 \, n'} C^{r_3 \, r_4 \, r''}_{m_3 \, m_4 \, m''}C^{r_3 \, r_4 \, r''}_{n_3 \, n_4 \, n''} \\
        & \times \int dU \; D^{r'}_{m' n'} D^{* r''}_{-m'' -n''}(-1)^{n''-m''}
    \end{align}
    \begin{align}
    \nonumber
        &= \sum_{r', m', n'} \sum_{r'', m'', n''} C^{r_1 \, r_2 \, r'}_{m_1 \, m_2 \, m'}C^{r_1 \, r_2 \, r'}_{n_1 \, n_2 \, n'} C^{r_3 \, r_4 \, r''}_{m_3 \, m_4 \, m''} \\
        & \times d_{r'}^{-1} (-1)^{n'' - m''} C^{r_3 \, r_4 \, r''}_{n_3 \, n_4 \, n''} \delta_{m', -m''}\delta_{n', -n''}\delta_{r',r''} \\
    \nonumber
        & = \sum_{r', m', n'} d_{r'}^{-1} (-1)^{m' - n'} \\
        & C^{r_1 \, r_2 \, r'}_{m_1 \, m_2 \, m'}C^{r_1 \, r_2 \, r'}_{n_1 \, n_2 \, n'} C^{r_3 \, r_4 \, r'}_{m_3 \, m_4 \, -m'} C^{r_3 \, r_4 \, r'}_{n_3 \, n_4 \, -n'}.
    \end{align}
    This allows us to write the partition function directly as
    
    \begin{align}
    \nonumber
        Z &= \sum_{\{r\text{'}s\}}\sum_{\{m\text{'}s\}}\sum_{\{n\text{'}s\}} \prod_{l} \left( F_{r_{1,l}}(\beta)F_{r_{2,l}}(\beta)F_{r_{3,l}}(\beta)F_{r_{4,l}}(\beta) \right)^{\frac{1}{2}} \\ 
    \nonumber
        & \times \sum_{r', m', n'} d_{r'}^{-1} (-1)^{m' - n'}  \\
        & C^{r_1 \, r_2 \, r'}_{m_1 \, m_2 \, m'}C^{r_1 \, r_2 \, r'}_{n_1 \, n_2 \, n'} C^{r_3 \, r_4 \, r'}_{m_3 \, m_4 \, -m'} C^{r_3 \, r_4 \, r'}_{n_3 \, n_4 \, -n'}.
    \end{align}
    and gives us a $T$ tensor of the form
    \begin{align}
    \nonumber
        & T_{(r_1, m_1, n_1)(r_2, m_2, n_2)(r_3, m_3, n_3)(r_4, m_4, n_4)} = \\
    \nonumber
        &\left( F_{r_{1}}(\beta) F_{r_{2}}(\beta) F_{r_{3}}(\beta) F_{r_{4}}(\beta) \right)^{\frac{1}{2}} \\
    \nonumber
        & \times \sum_{r', m', n'} d_{r'}^{-1} (-1)^{m' - n'} \\
        & C^{r_1 \, r_2 \, r'}_{m_1 \, m_2 \, m'}C^{r_1 \, r_2 \, r'}_{n_1 \, n_2 \, n'} C^{r_3 \, r_4 \, r'}_{m_3 \, m_4 \, -m'} C^{r_3 \, r_4 \, r'}_{n_3 \, n_4 \, -n'}.
    \end{align}
    This $T$ tensor can be used just as a typical spin-model tensor with four (grouped) indices.  The typical contraction between tensor legs can be carried out with the help of a grouped set of Kronecker deltas.
    \begin{equation}
        \tilde{\delta}_{(r, i, i')(r', j, j')} = \delta_{r, r'}\delta_{i,j}\delta_{i', j'}.
    \end{equation}
    This tensor ensures the same representation per link, and circulates the trace of the matrix indices along the link between sites.

\section{TRG Formulations of Lattice Gauge Models}
    \label{sec:gauge}
    In this section, the tensor-network forms for the partition function of Abelian gauge models including 3D $Z_2$ gauge theory and D-dimensional $U(1)$ gauge models (D $=2,3,4$) are shown. Two formulations of the TRG method are constructed: one is left-right asymmetric and the other symmetric.    
    \subsection{Three-dimensional $Z_2$ Gauge Theory} 
\begin{figure}[hb]
            \begin{center}
            	\includegraphics[width=7cm]{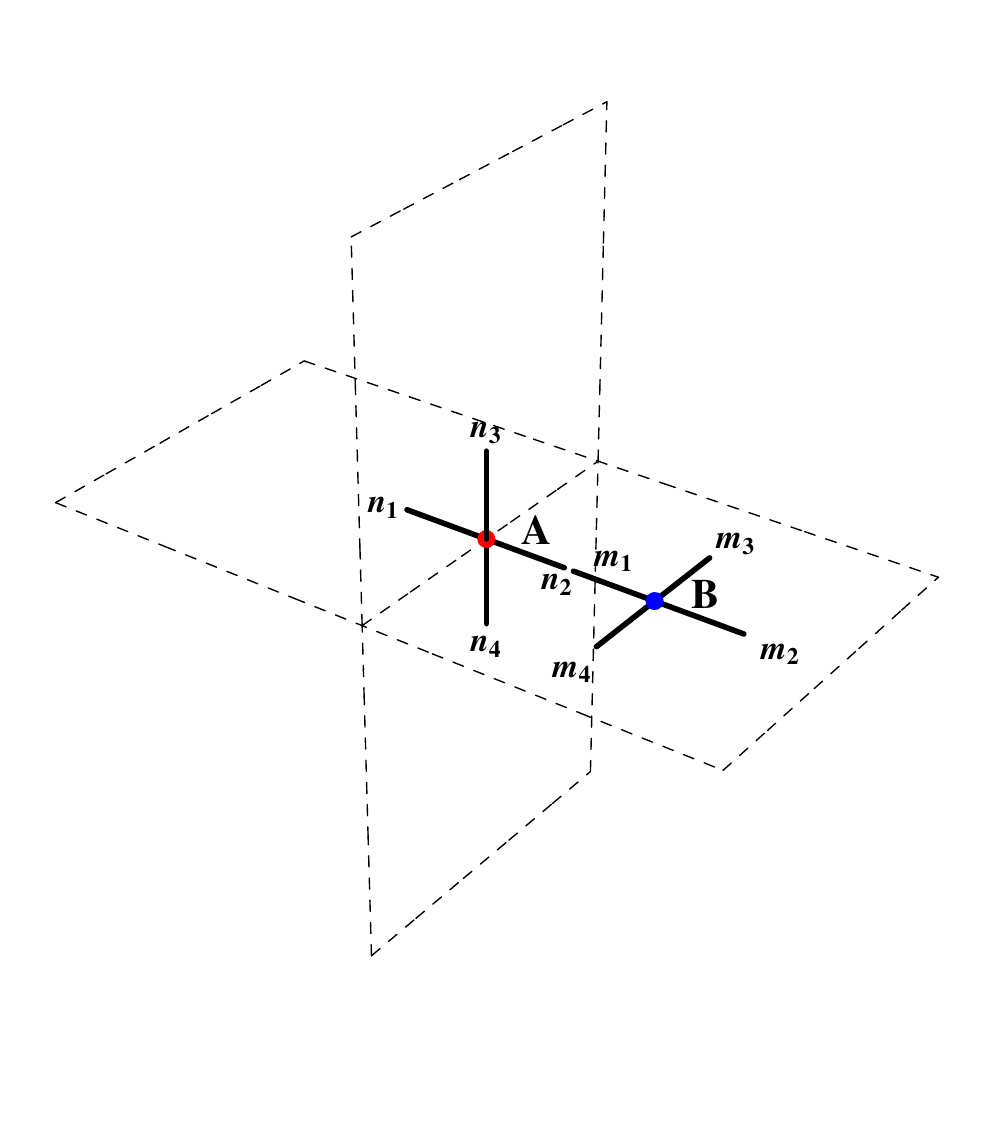}
            \end{center}
            \caption{\label{fig:abtr} $A$ tensor and $B$ tensors}
        \end{figure}

        We first consider a simple gauge theory on a lattice, the three-dimensional compact $Z_2$ gauge theory, with the partition function,
        \begin{equation}
            Z=\sum_{\{\sigma\}}\exp\left(\beta\sum_{P}\sigma_{12}\sigma_{23}\sigma_{34}\sigma_{41}\right),
        \end{equation}
        where the action is a sum over all the plaquettes and the field $\sigma_{ij}=\pm 1$ are attached to each link of the lattice. We can now proceed as in Eq. (\ref{eq:char}) and write 
        a single plaquette contribution using a sum with $n=0$ or 1 of
         \beq(\sqrt[4]{\tanh(\beta)}  \sigma_{12} \sqrt[4]{\tanh(\beta)}  \sigma_{23} \sqrt[4]{\tanh(\beta)}
       \sigma_{34} \sqrt[4]{\tanh(\beta)}  \sigma_{41})^{n}.\nonumber\enq
        Regrouping the factors with a given $\sigma_l$ and summing over $\pm 1$ we obtain a tensor attached to this link
         \begin{eqnarray}
            \label{eq:ttensor}
            \nonumber
            A^{(l)} _{n_1n_2n_3n_4}=&&\left(\sqrt[4]{\tanh\beta}\right)^{n_1+n_2+n_3+n_4}\times \\
            &&\delta\left(\text{mod}[n_1+n_2+n_3+n_4,2]\right).\\ \nonumber
        \end{eqnarray}
        The four links attached to a given plaquette $p$ must carry the same index 0 or 1. For this purpose we introduce a new tensor 
         \begin{eqnarray}
            \label{eq:ttensorB}
            \nonumber
             B^{(p)}_{m_1m_2m_3m_4}=&&\delta(m_1,m_2,m_3,m_4)\\
            =&&\begin{cases} 
                1,  &\mbox{all } n_i\mbox{ are the same} \\
                0, & \mbox{otherwise}.
            \end{cases}
        \end{eqnarray}
        The partition function can now be written as
        \begin{equation}
            Z=(2\cosh\beta)^{3V} \Tr \prod_{l}A^{(l)}_{n_1n_2n_3n_4}\prod_{p}B^{(p)}_{m_1m_2m_3m_4},
        \end{equation}
        where $V$ is the volume of the system and $\Tr$ is a notation for sum over all the shared plaquettes. A graphical representation of the tensors in 
        provided in Fig. \ref{fig:abtr}.
        One can check that the new expression for the partition function reproduces the strong coupling expansion.
         \subsubsection{Asymmetric Formulation}

        \begin{figure}[h]
                \begin{center}
                	\includegraphics[width=6cm]{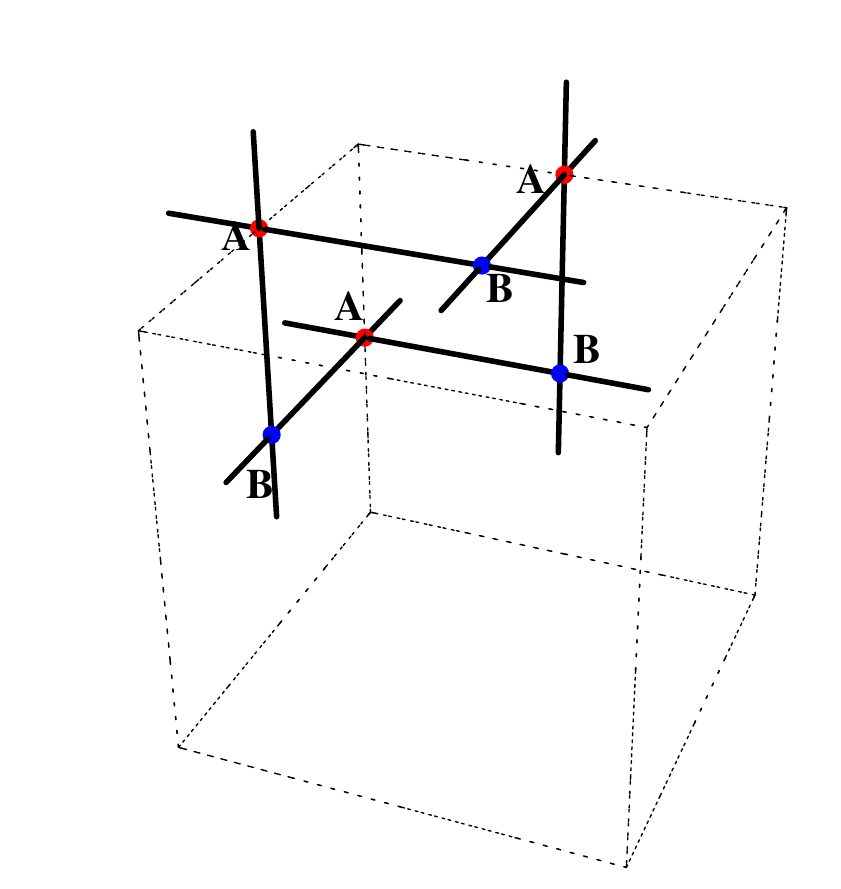}
                	\includegraphics[width=6cm]{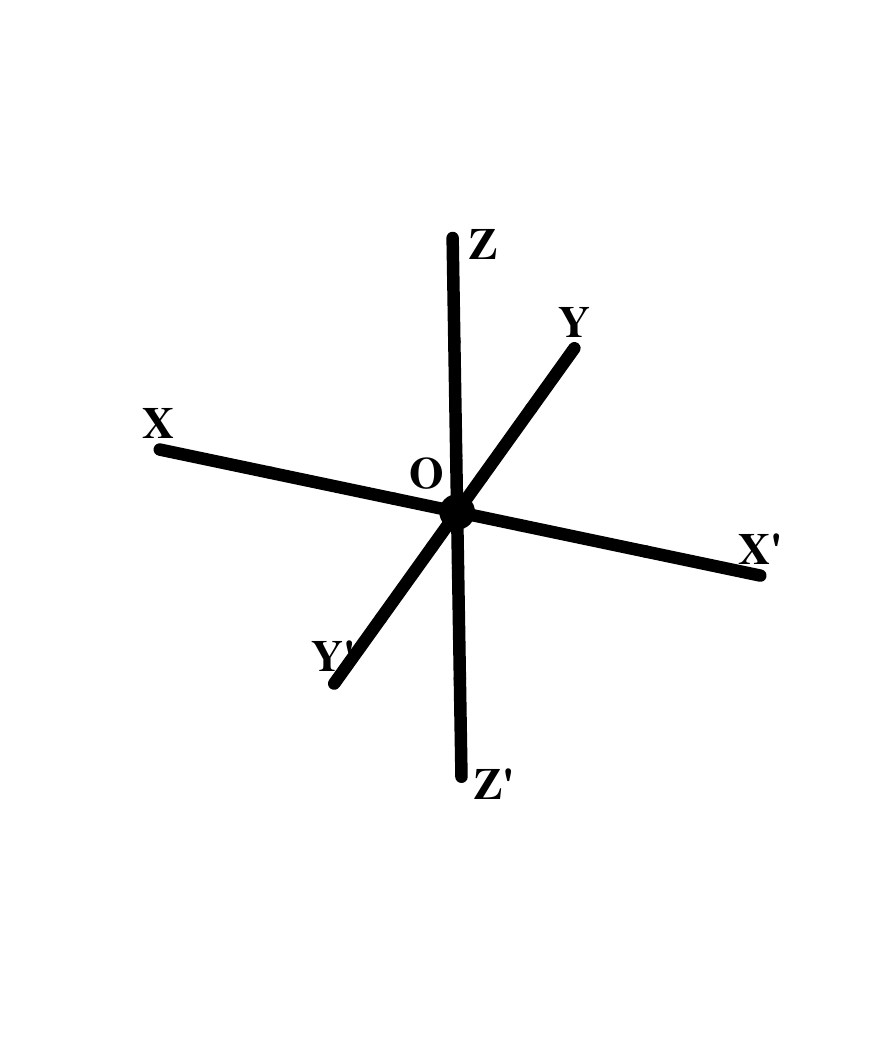}
                \end{center}
                \caption{\label{fig:t3d0} (top) A new basic cell in an original cube. The equivalent $T_6$ tensor  (bottom), its center is (1/4, 3/4, 3/4) in the original cube.}
            \end{figure}

            By using 3 $A$ tensors and 3 $B$ tensors as shown Fig.\ref{fig:t3d0}, a basic cell can be constructed. There are twelve external legs. We can recombine the indices attached to the legs pointing in the same directions using product states (labeled by capital letters). For instance $X=x_1\otimes x_2$ and similarly with the other directions. 
            Proceeding this way, we obtain a new tensor ${T_6}_{XX'YY'ZZ'}$ which can be treated as in the case of a 3D spin model. However, in the 
            positive ($X, Y, Z$) and negative ($X', Y', Z'$) directions, the opposite legs are associated with different tensors. For instance $X$ is associated with $A$ and $X'$ with $B$.             

                        The partition function can be rewritten as the tensor-network state of the new $T_6$ tensor at each cube $c$,
            \begin{equation}
                Z=(2\cosh\beta)^{3V}\text{Tr}\prod_{c}{T_6}^{(c)}_{XX'YY'ZZ'} \ .
            \end{equation}
 To blockspin, we can use anisotropic steps by contracting the lattice alternatively in the x axis, y axis, and z axis directions. In each step, the lattice size is reduced by a factor of 2 
   in the  appropriate direction and a new $T_6'$ tensor is generated as,
            \begin{eqnarray}
                \label{eq:t6p}
                \nonumber
                &&{T_6'}_{XX''\tilde{Y}(Y_1,Y_2)\tilde{Y'}(Y_1',Y_2')\tilde{Z}(Z_1,Z_2)\tilde{Z'}(Z_1',Z_2')}\\
                &&=\sum_{X'}{T_6}_{XX'Y_1Y_1'Z_1Z_1'}{T_6}_{X'X''Y_2Y_2'Z_2Z_2'},
            \end{eqnarray}
            where $\tilde{Y}(Y_1,Y_2)$ is the notation for the product states $\tilde{Y}=Y_1\otimes Y_2$ and similarly with the other directions. The partition function can then be rewritten 
            as the trace of product of $T_6'$ tensors as before blocking. 
            
            It is straightforward but tedious to write an isotropic blocking formula involving the product of 8 $T_6$tensors. It is also possible to find tensors associated with the partition function in 
            the temporal gauge. The $A$ tensor on the temporal links disappear while those on the space links have a space-time asymmetry. This will be important for numerical applications. 

        \subsubsection{Symmetric Formulation}
\begin{figure}[h]
                \begin{center}
                    \includegraphics[width=7cm]{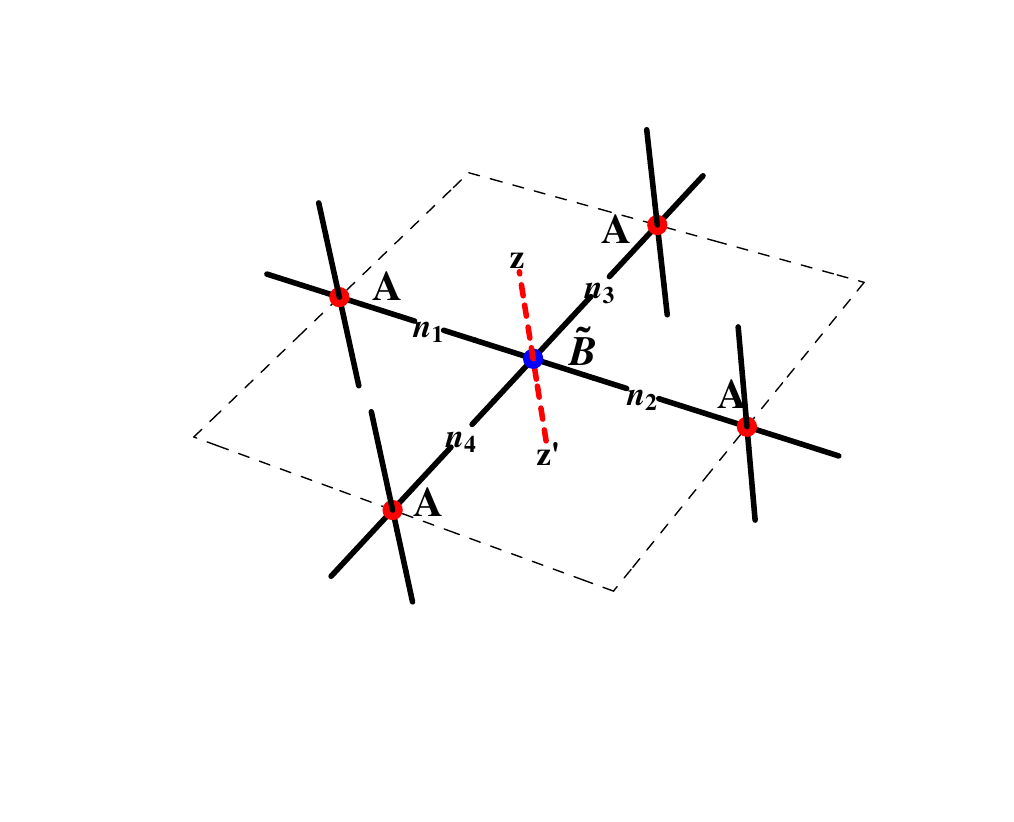}
                \end{center}
                \caption{\label{fig:isot1} $A$ and $B'$ tensor on each plaquette.}
            \end{figure}

\begin{figure}[h]
                \begin{center}
                    \includegraphics[width=7cm]{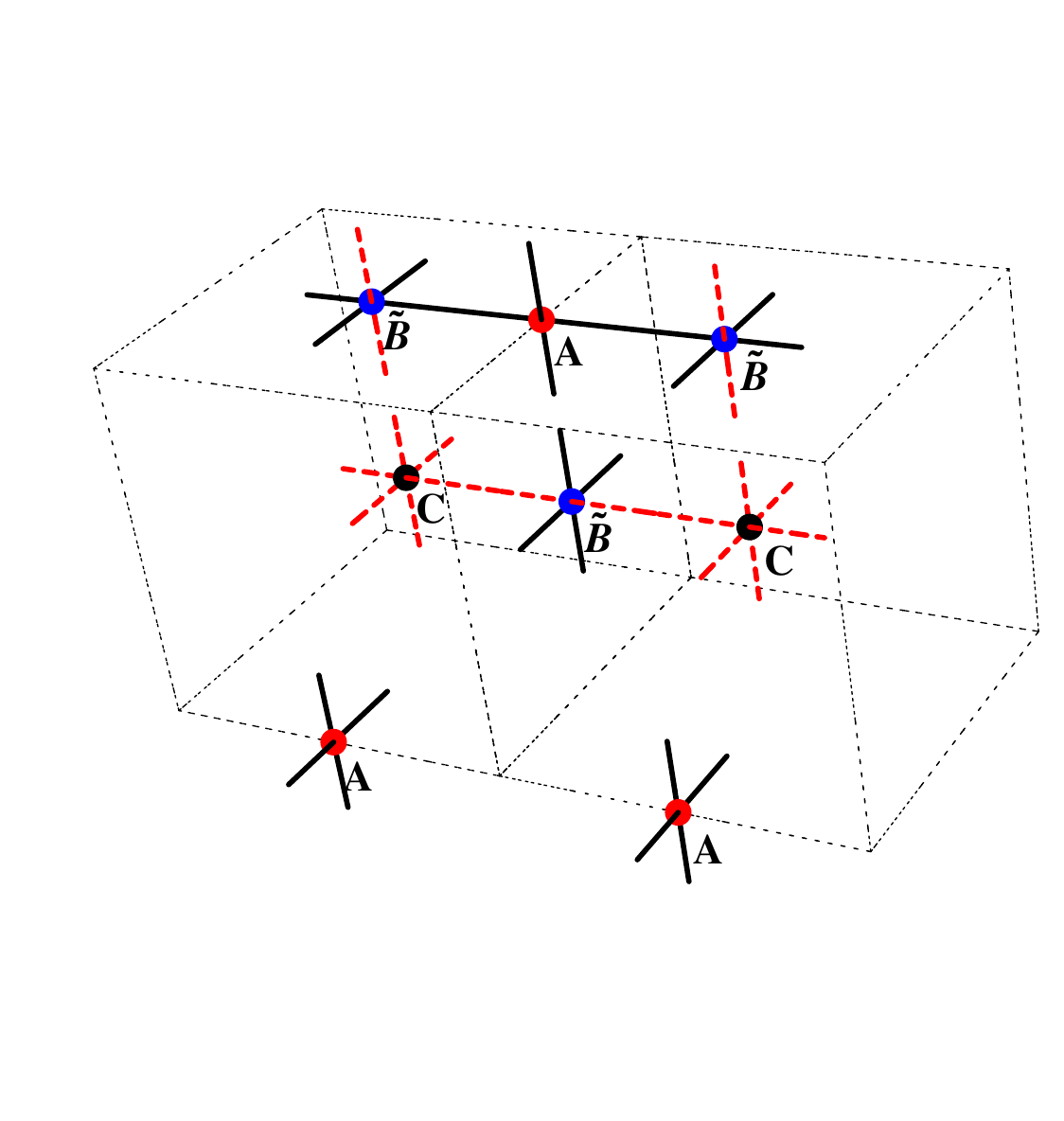}
                \end{center}
                \caption{\label{fig:isotc} blocking procedure}
            \end{figure}

        The difference between the positive and negative directions in the previous formulation can be avoided by introducing new tensors. First, we notice that the $A$ and $B$ tensors do not suffer from this asymmetry. However they do not close under blocking. To see this we can try to combine the $B$ tensors of two adjacent plaquettes in the same plane into a new one. This does not work because the $A$ tensor on the common link induces two new legs orthogonal to the plane and pointing in opposite directions. This is the effect that is eliminated in the Migdal-Kadanoff approximation by bond-sliding. Here, we want an exact formula so we modify the $B$ tensor to form a $\tilde{B}$ tensor with 6 indices (see Fig. \ref{fig:isot1}) with initial value 
            \begin{equation}
                \label{eq:bptensor}
                \tilde{B}_{n_1n_2n_3n_4zz'}=B_{n_1n_2n_3n_4}\delta_{zz'} \ ,
            \end{equation}
          for a plaquette in the $x-y$ plane and with similar expressions for the two other planes. The new legs piercing the plaquettes can be traced by introducing a new 
   tensor $C_{xx'yy'zz'}$ at the center of the cubes, as shown in Fig. \ref{fig:isotc}, with initial value 
            \begin{equation}
                \label{eq:ctensor}
                C_{xx'yy'zz'}=\delta_{xx'}\delta_{yy'}\delta_{zz'},
            \end{equation}
            where the $\delta_{ij}$ is the Kronecker delta function. In general, the $C$ tensor has its indices as the $T_6$ tensor shown at the bottom of Fig. \ref{fig:abtr}, but its center is (1/2, 1/2, 1/2) in the original cube.  
            
            We can now rewrite the partition function as 
             \begin{equation}
            Z=K(2\cosh\beta)^{3V} \Tr \prod_{l}A^{(l)}\prod_{p}\tilde{B}^{( p )}\prod_{c}C^{( c )} \ ,
        \end{equation}
          where the indices are implicit to keep the formula short. The Kronecker delta in the initial values can be summed along open or closed lines 
          (depending on the boundary conditions) and give rise to a power of 2 that can be eliminated by adjusting the constant $K$. The other traces are as in the original expression of 
          the partition function. 
   
            A blocking procedure can be constructed by sequentially combining two cubes into one in each of the directions. This is illustrated in one direction in Fig. \ref{fig:isotc},      
            
            We can write explicit blocking formulas. 
            On the link of the new lattice formed by two cubes, two parallel $A$ tensors form the new $A'$ tensor with product states (capital letters). Each tensor element is
            \begin{eqnarray}
                \label{eq:newa1}
                \nonumber
                &&{A'}_{X(x_1,x_2)X'(x_1',x_2')Y(y_1,y_2)Y'(y_1',y_2')}\\
                &&=A_{x_1x_1'y_1y_1'}\times A_{x_2x_2'y_2y_2'}.
            \end{eqnarray}
            On the new face, two $\tilde{B}$ tensors and one $A$ tensor form a new $\tilde{B}'$ tensor,
            \begin{eqnarray}
                \label{eq:newb1}
                \nonumber
                &&{\tilde{B}'}_{xx'Y(y_1,y_2)Y'(y_1',y_2')Z(z_1,z_2,z_3)Z'(z_1',z_2',z_3')}\\
                &&=\sum_{x_3,x_3'}\tilde{B}_{xx_3y_1y_1'z_1z_1'}A_{x_3x_3'z_3z_3'}\tilde{B}_{x_3'x'y_2y_2'z_2z_2'}.
            \end{eqnarray}
            At the center, two $C$ tensors and one $\tilde{B}$ tensor form a new $C'$ tensor,
            \begin{eqnarray}
                \label{eq:newc1}
                \nonumber
                &&{C'}_{xx'Y(y_1,y_2,y_3)Y'(y_1',y_2',y_3')Z(z_1,z_2,z_3)Z'(z_1',z_2',z_3')}\\
                &&=\sum_{x_2,x_2'}C_{xx_2y_1y_1'z_1z_1'}\tilde{B}_{x_2x_2'y_2y_2'z_2z_2'}C_{x_2'x'y_3y_3'z_3z_3'}.
            \end{eqnarray}

    \subsection{$U(1)$ Gauge Models}
        In this section, we formulate the $U(1)$ gauge models in $D$ dimensions in terms of tensor-network states. The partition function of these models can be written as 
        \begin{equation}
            Z=\prod_{\langle ij\rangle}\int_{-\pi}^{\pi}\frac{d\theta_{ij}}{2\pi}\exp\left(\beta\sum_{P}
            \cos(\theta_{12}+\theta_{23}-\theta_{43}-\theta_{14})\right),
        \end{equation}
        where the product is running through all the links of the lattice and the sum is over all the plaquettes.

        Using the Fourier expansion with the Bessel functions as in Eq. (\ref{eq:bessel}) and collecting the factors for each link, we obtain the tensor 
         \beq
            \label{eq:tndu1}
            A_{n_1\dots n_{2(D-1)}}=\prod_{i=1}^{2(D-1)} \sqrt[4]{I_{n_i}(\beta)}
            \delta\left(\sum_{i=1}^{2(D-1)}(-1)^{i+1}n_i\right)   ,     \enq 
            where the $I_n$s are the modified Bessel functions. For any $D$, we can use a $B$ tensor that ensures that the four indices attached to a
            plaquette are identical just like for the $Z_2$ case. The partition function can be written as
        \begin{equation}
            Z=\text{Tr}\prod_{l}A^{(l)}_{n_1\dots n_{2(D-1)}}\prod_{p}B^{(p)}_{m_1m_2m_3m_4}.
        \end{equation}
     We can construct the blocking procedure in both the asymetric way and symmetric way following what has been done for $Z_2$.  From a geometric viewpoint, a basic cell in a $D$ dimensional lattice contains $D$ $A$ tensors with $2(D-1)$ legs each and $\small{\frac{D(D-1)}{2}}$ $B$ tensors, each always with four legs. We now will consider $D=2$, 3 and 4 separately.

        \subsubsection{$D=2$}
\begin{figure}[h]%
                \begin{center}
                    \includegraphics[width=6cm]{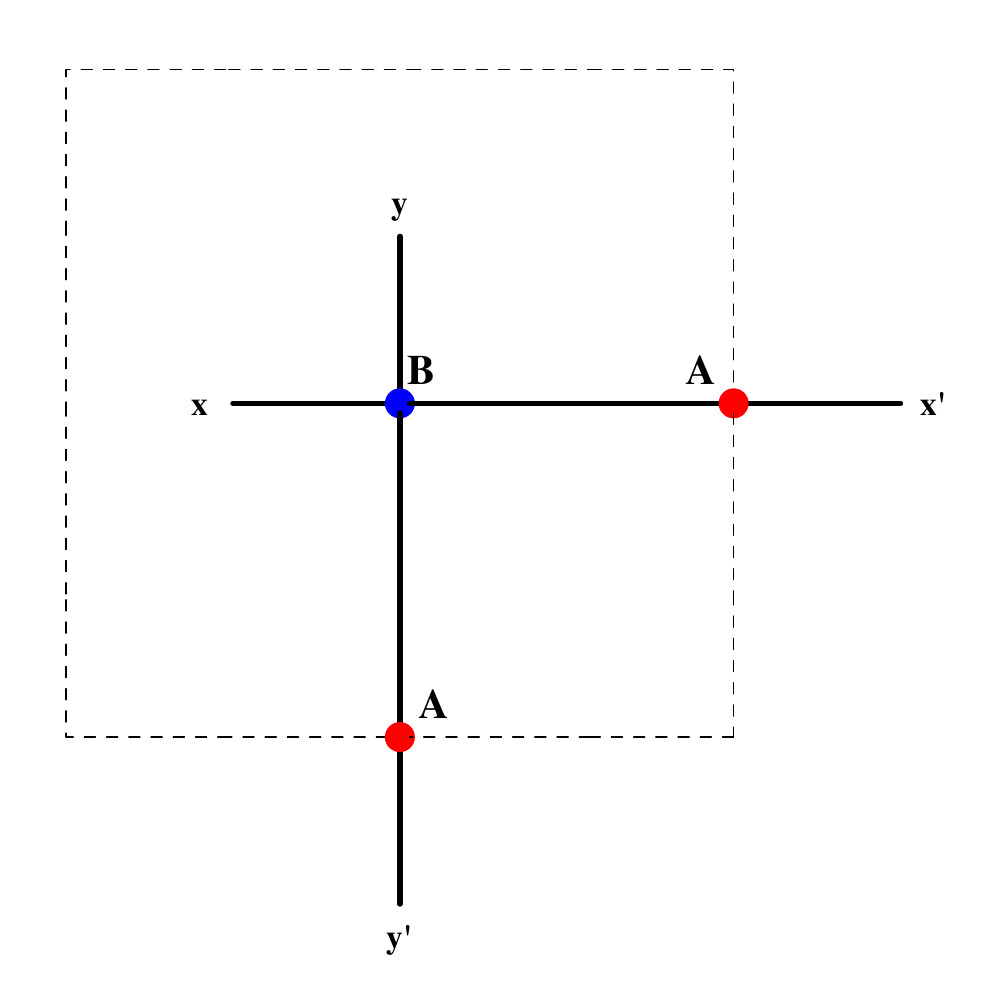}%
                    \caption{$T_4$ tensor which contains two $A$ tensors and one $B$ tensor. }%
                \label{fig:tdt4t}%
                \end{center}
            \end{figure}

        For $D$= 2, the $A$ tensor is just proportional to a Kronecker delta. This allows us to block two adjacent $B$ tensors and get another $B$ tensor. 
        We can also construct an asymmetric tensor $T_4$ from a basic cell as illustrated in 
        Fig. \ref{fig:tdt4t}. If one leg is fixed, all the other three are also fixed because of the constraint of the $B$ tensor. Thus, 
            \begin{equation}
                T_4\equiv T_{xx'yy'}=I_x(\beta)\delta(x,x',y,y')
            \end{equation}
                This tensor can be blocked isotropically with result     $(I_x(\beta))^4\delta(x,x',y,y')$. For periodic boundary conditions we can pursue this process 
                and we get the known answer    
                  \begin{equation}
                Z=\sum_{n=-\infty}^{\infty}I_n(\beta)^{L_x\times L_y}.
            \end{equation}
            where $L_x\times L_y$ is the area of the system. For open boundary conditions, we can represent the three 
            indices tensor at the boundary as a a four indices tensor with an index 0 for the leg going outside the boundary. With this, only the $n=0$ term survives from the sum obtained with periodic boundary conditions. This is very similar to the 1D $O(2)$ model. 

        \subsubsection{$D=3$}
        The treatment is almost identical to the 3D 
           $Z_2$ gauge model. The geometric construction is the same but the initial tensor is given by Eq. (\ref{eq:tndu1}) and the initial sums run over the integers instead of 0 and 1 for $Z_2$. 
                   \subsubsection{$D=4$}
\begin{figure}[h]
                \begin{center}
                    \includegraphics[width=6.5cm]{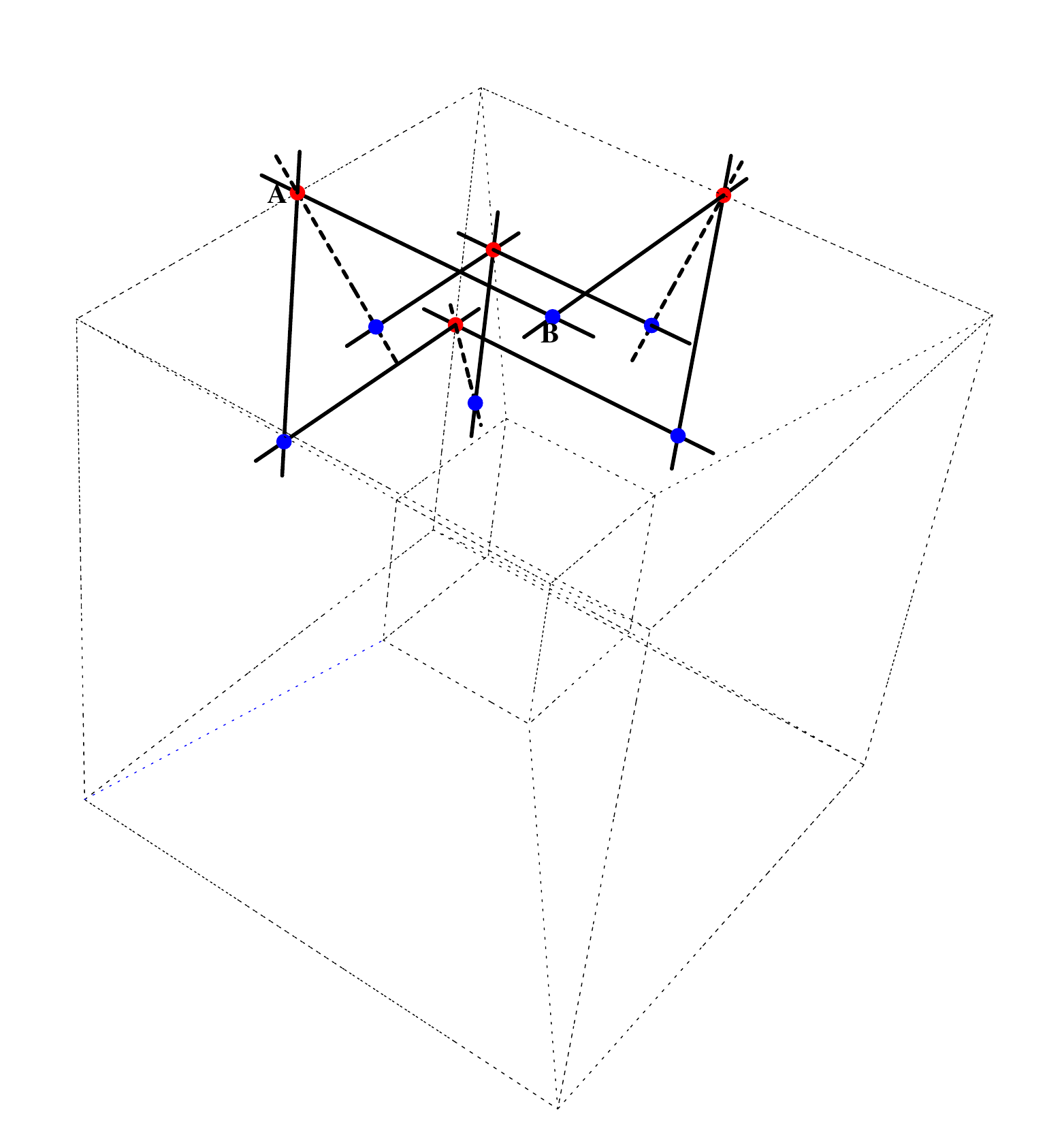}%
                    \caption{\label{fig:tcube} 4-D $U(1)$ tensor in a hypercube red (online) dots: $A$ tensors. blue (online) dots: $B$ tensors.}%
                \end{center}
            \end{figure}

            The basic cell of tensors in $D=4$ is illustrated in  Fig. \ref{fig:tcube}). There are four $A$ tensors with six legs and six $B$ tensors in one basic cell of the hyper-cube. There are 3 legs pointing in each of the directions. 
            Following the asymmetric procedure, we can combine each of these three legs into a single index, build a rank 8 tensor and block as in the spin model case. It seems possible to follow the symmetric procedure and build a modified $B$ tensor with 4 additional legs in the directions orthogonal to the plaquette, a modified $C$ tensor with 2 additional legs pointing in the direction orthogonal to the cubes and a new tensor with 8 legs located at the center of the hypercubes. Blocking in one direction is then performed by contracting two similar tensors with a tensor associated to the object with one less dimension in between. 

    \subsection{Abelian factorization}
   All the initial $A$ tensors calculated have the factorization property shown by Abelian spin models and discussed in Sec. \ref{subsec:factor}. The same reasoning can be used for 
   Abelian gauge theories. Assuming the character expansion for the single plaquette weight
        \beq
            {\rm e}^{-\beta S_{p}} = \sum_{r} F_{r}(\beta) \chi^{r}(U)\ ,
        \enq
we can rewrite  $ \chi^{r}(U)$ as the product of the 4 characters for each of the 4 links $\chi^{r}(U_1)\chi^{r}(U_2)\chi^{r}(U_3)\chi^{r}(U_4)$,  
factorize the contributions associated with each link and then integrate over the link variables. 
The initial tensor reads: \beq
            A_{ijkl} = \left( F_{i}(\beta) F_{j}(\beta) F_{k}(\beta) F_{l}(\beta) \right)^{\frac{1}{4}} \delta_{i+k,j+l}^{g}.
        \enq
 The main difference with the spin model is the appearance of the fourth root instead of the square root. 
\subsection{TRG Formulation of 3D $SU(2)$ Gauge}
        Using the conventions from Ref. \cite{Varshalovich:1988ye}, and following a procedure described in Ref. \cite{Anishetty:1992xa}, we start with a partition function for the 3D $SU(2)$ gauge model
        \begin{equation}
            Z = \prod_{ni} \int dU(ni) \prod_{nij} \exp{\left\{ \frac{\beta}{4} {\rm Re} [ \tr \left[ U(nij) \right] ] \right\}},
        \end{equation}
        with $U(nij)$ the product of group elements around a plaquette and $ni$ the links of the lattice.  Since the action only depends on the trace of matrix representations of $SU(2)$, we can re-write the action as a character expansion
        \begin{equation}
            {\rm e}^{-\beta S_{p}} = \sum_{r} F_{r}(\beta) \chi^{r}(U(nij)).
        \end{equation}
        Then the partition function can be written as
        \begin{equation}
            Z = \prod_{ni} \int dU(ni) \prod_{nij} \sum_{r(nij)} F_{r(nij)}(\beta) \chi^{r(nij)}(U(nij)).
        \end{equation}
        We can re-write the characters of the product of group elements as the trace over the product of the matrix representations of group elements
        \begin{equation}
            \chi^{r}(U_1 U_2 U_3 U_4) = D^{r}_{ij}(U_1) D^{r}_{jk}(U_2) D^{r}_{kl}(U_3) D^{r}_{li}(U_4),
        \end{equation}
        and with these ``Wigner D-functions'' we can perform the product over plaquettes of the lattice, and gather together the four D-functions which all share the same link variable.  In 3D there are four plaquettes for each link and so there are four D-functions per link variable.  This situation is identical to the 2D Principal Chiral model, since in 2D, there are four links impinging on a site.  The only minor difference in this case, is that each plaquette is bordered by four links, as opposed to 2D where each link is bordered by two sites.  The consequence of this is that the character coefficients are shared more between the links.  We can use the same analysis as before for the integration and write down the partition function directly.   

        \begin{align}
        \nonumber
            Z &= \prod_{ni} \sum_{r,m,n} \int dU(ni) \left( F_{r_{1}}(\beta) F_{r_{2}}(\beta) F_{r_{3}}(\beta) F_{r_{4}}(\beta) \right)^{\frac{1}{4}} \\
            &\times D^{r_1}_{m_1 n_1}(U) D^{r_2}_{m_2 n_2}(U) D^{r_3}_{m_3 n_3}(U) D^{r_4}_{m_4 n_4}(U).
        \end{align}
        \begin{align}
        \nonumber
            &= \prod_{ni} \sum_{r\text{'s}} \sum_{m\text{'s},n\text{'s}} \left( F_{r_{1}}(\beta) F_{r_{2}}(\beta) F_{r_{3}}(\beta) F_{r_{4}}(\beta) \right)^{\frac{1}{4}} \\
        \nonumber
            &\times \sum_{r', m', n'} d_{r'}^{-1} (-1)^{m' - n'} \\
            & C^{r_1 \, r_2 \, r'}_{m_1 \, m_2 \, m'} C^{r_1 \, r_2 \, r'}_{n_1 \, n_2 \, n'} C^{r_3 \, r_4 \, r'}_{m_3 \, m_4 \, -m'} C^{r_3 \, r_4 \, r'}_{n_3 \, n_4 \, -n'}.
        \end{align}
        This gives us an $A$ tensor of the form
        \begin{align}
        \nonumber
            & A_{(r_1, m_1, n_1)(r_2, m_2, n_2)(r_3, m_3, n_3)(r_4, m_4, n_4)} = \\
        \nonumber
            &\left( F_{r_{1}}(\beta) F_{r_{2}}(\beta) F_{r_{3}}(\beta) F_{r_{4}}(\beta) \right)^{\frac{1}{4}} \\
        \nonumber
            & \times \sum_{r', m', n'} d_{r'}^{-1} (-1)^{m' - n'} \\
            & C^{r_1 \, r_2 \, r'}_{m_1 \, m_2 \, m'}C^{r_1 \, r_2 \, r'}_{n_1 \, n_2 \, n'} C^{r_3 \, r_4 \, r'}_{m_3 \, m_4 \, -m'} C^{r_3 \, r_4 \, r'}_{n_3 \, n_4 \, -n'}.
        \end{align}
        Now, the model demands that there be a single representation assigned to each plaquette, we see this during the character decomposition when a single plaquette takes on a single representation.  However, while the D-function matrix indices are traced out, they demand to be traced out in a specific way, namely, to close around a plaquette.  This behavior must be obeyed during the tensor reconstruction.  This can be handled by separate tensors.  One tensor, the original $B$ tensor from Abelian models can remain the same and is responsible for keeping the representations the same on a plaquette.  Next we need adjacent (link) matrix indices, $m$s and $n$s, to be contracted.  This is achieved with four Kronecker deltas each contracting a pair of adjacent indices. The initial value of the tensor is given by
        \begin{eqnarray}
          &&  \tilde{B}_{(r, i, i')(r', j, j')(r'', k, k')(r''', l, l')} \nonumber \\ &&=B_{r r' r'' r'''}\delta_{i, j}\delta_{j', k} \delta_{k', l} \delta_{l', i'} \ .
        \end{eqnarray}
       We can now proceed as in the 3D Abelian case to write the partition function and perform blockings using $A$ and $B$ tensors. The only difference is that the  single indices of the Abelian formulas need to be replaced by three indices. 
\section{Conclusions}
In conclusions, we have shown that the partition functions of the 2D  $O(2)$ and $O(3)$ sigma models, the 2D $SU(2)$ principal chiral model and for the 3D gauge theories with group $Z_2$, $U(1)$ and $SU(2)$ can be written in terms of local tensors and that exact blocking formulas can be written for these models. 
The basic ingredient is the character expansion. This is available for any finite or compact group. For Abelian models, the factorization properties discussed for spin and gauge models should guarantee that the procedures described here can be extended for any compact Abelian group in any dimension. In the non-Abelian case, it is in addition necessary to 
reexpress products of representations in terms of irreducible representations. Again this is possible for $O(N)$ and $SU(N)$ with larger $N$ than the ones considered here.  

Models with fermions have not been discussed. In Ref. \cite{PhysRevB.87.064422}, it was shown that standard SVD methods can be used to factorize exponentials of quadratic forms in Grassman numbers and then perform the local integrations.
This yields tensors similar to the ones constructed for spin models. Combining this result with the ones for the gauge models presented here is an important goal. A first objective could be the 2D Schwinger model. 
It is interesting that the quantum treatment of this model in 1+1 dimensions in terms of tensor network states has been proposed recently \cite{Banuls:2013jaa}. It would be very interesting to understand the standard quantum-classical correspondence in a unified tensor language.

Exact blocking formulas maybe useful for analytical problems such as the understanding of confinement for 4D $SU(2)$ and the lack thereof for $U(1)$ gauge theories \cite{Tomboulis:2009zz}.
There are many possible numerical applications of the blocking formulas presented here. The finite size of computer memory requires truncations and projections which are model dependent. 
The 2D $O(2)$ model can be treated as the 2D Ising model. The good agreement between TRG and Monte Carlo calculations of thermodynamics quantities and critical properties will be reported elsewhere 
\cite{groupprogress}. Numerical implementations for 3D Ising gauge theory are under progress. 

In general, the computational demands are very different from those present in Monte Carlo simulations. For a given set of states in the external legs (which communicate with the other blocks as in  Fig. \ref{fig:square}), the 
sums over the internal states amounts to solve a small lattice problem and takes little CPU time. However, the large number of combinations of external states requires many repetitions and can be stretching the limit of computer memory. 

The numerical treatment seems insensitive to sign problems. In conventional Monte Carlo simulations, calculations with complex $\beta$ can only be achieved by reweighing of configurations obtained with real $\beta$ where the sign problem is absent. This only allows small imaginary values of $\beta$. In contrast, the TRG method allows larger imaginary parts. This allowed us to calculate the zeros of the partition function for the 2D Ising and $O(2)$ models in good agreement with existing results \cite{groupprogress}. We plan to use the TRG method to study the $Z_N$ clock and $O(2)$ models with a complex chemical potential and 
compare the results with those obtained with dual formulations \cite{Meisinger:2013zfa} and world-line methods \cite{Banerjee:2010kc,PhysRevLett.87.160601}. 

In summary, the TRG method is a very promising method to deal with models studied by lattice gauge theorists. We hope that the recent numerical success will extend to other models and that ultimately it will be useful to approach important problems such as the phase diagram of QCD and the boundary of the conformal window for various multiflavor gauge theories. 

\begin{acknowledgments}

This research was supported in part  by the Department of Energy
under Award Numbers DE-SC0010114 and FG02-91ER40664. 
Preliminary numerical work checking the validity of analytical formulas presented here used resources of the National Energy Research Scientific Computing Center, which is supported by the Office of Science of the U.S. Department of Energy under Contract No. DE-AC02-05CH11231. 
Y. L. is supported by the URA Visiting Scholars' program. 
Fermilab is operated by Fermi Research Alliance, LLC, under Contract
No.~DE-AC02-07CH11359 with the United States Department of Energy.
 Our work on the subject started  while attending the KITPC workshop ``Critical Properties of Lattice Models" in summer 2012. 
 Y. M. did part of the work while at the workshop ``LGT in the LHC Era" in summer 2013 at the Aspen Center for Physics supported by NSF grant No 1066293. 
 We thank  M.C. Banuls, S. Chandrasekharan, A. Denbleyker, A. Hasenfratz, Anyi Li, M. Ogilvie, P. Orland, W. Polyzou, C. Pryor, V. Rodgers,  T. Tomboulis, and  X-G Wen,  for valuable conversations and suggestions. 
\end{acknowledgments}

%



\end{document}